\documentclass[twocolumn,showpacs,preprintnumbers,amsmath,amssymbm,prl]{revtex4}

\usepackage{graphicx}
\usepackage{dcolumn}
\usepackage{bm}
\usepackage{epsf}
\usepackage{amsfonts}

\newcommand{\eg}{\emph{e.g., }}
\newcommand{\beq}{\begin{equation}}
\newcommand{\eeq}{\end{equation}}
\newcommand{\beqr}{\begin{eqnarray} \nonumber}
\newcommand{\eeqr}{\end{eqnarray}}
\newcommand{\beqrb}{\begin{eqnarray}}
\newcommand{\eeqrb}{\nonumber \end{eqnarray}}

\newcommand{\fin}{\mbox{ .}}
\newcommand{\coma}{\mbox{ ,}}

\newcommand{\lrgspc}{\,\,\,\,\,\,\,\,\,}

\newcommand{\myA}{{\mathcal{A}}}
\newcommand{\myEnt}{{\mathcal{S}}}
\newcommand{\myR}{{R}}
\newcommand{\myS}{{S}}

\newcommand{\nI}{\mathcal{I}}
\newcommand{\nT}{\mathcal{T}}
\newcommand{\nR}{\mathcal{R}}

\begin{document}

\title{The missing asymptotic sector of rotating black-hole spectroscopy}

\author{Uri Keshet}

\email{ukeshet@bgu.ac.il}

\author{Arnon Ben-Meir}

\affiliation{Physics Department, Ben-Gurion University of the Negev, Be'er-Sheva 84105, Israel}

\date{\today}

\begin{abstract}
The rotation of a black hole splits its spectrum in two, yet only one sector is known in the highly-damped regime.
We find the second, at least partly oblate sector, with quasinormal modes approaching the total reflection frequencies $\omega(n\gg1)\simeq m\Omega-2\pi i T (n-s)$, where $\Omega$ and $T$ are the horizon angular velocity and temperature, $s$ is the field spin, and $m$ is an azimuthal eigenvalue.
Some physical implications are discussed.
\end{abstract}


\maketitle

The natural vibrational modes of a black hole, known as quasinormal modes (QNMs) due to their typically fast damping, play an important role in the classical study of black holes, in the search for a theory of quantum gravity, and in the gauge/gravity duality; see \cite{BertiEtAl09, KonoplyaZhidenko11} for reviews.
The highly-damped modes, in particular, tend to show an interesting structure with simple or no dependence upon the vibration parameters.
It was suggested that such an asymptotic QNM corresponds directly to a quantum transition, and for a Schwarzschild black hole may be interpreted as an elementary change in the black hole area \cite{Hod98}.
The subsequent flurry of black hole spectroscopy research produced many results, but for the most part, the asymptotic QNMs defied a simple interpretation along the lines of the correspondence principle \cite{BertiEtAl09}.

The asymptotic QNMs of rotating black holes were computed numerically \cite{Berti04} and analytically \cite{KeshetHod07}.
These, and the related total reflection and total transmission modes, show a curious structure, which provides interesting hints at the microscopic description of rotating black holes in the asymptotic regime \cite{KeshetNeitzke08}.
Curiously, while the rotation is known to split the spectrum into two distinct branches of QNMs at low frequencies \cite{Leaver85}, only one asymptotic branch has been reported thus far, and was identified with equatorial vibrations \cite{KeshetNeitzke08}.
It is natural to ask what happens at asymptotic frequencies to the second QNM branch, and to polar excitations in general.
The question is further motivated by some contradictory results in the literature, and because the very existence of a second asymptotic sector was largely overlooked.

We compute the QNMs of Kerr black holes up to high overtones, in order to uncover the fate of the elusive second branch.
The transmission-reflection problem is then analyzed in the oblate limit, relevant to the new sector.

\emph{Teukolsky's equations.--- }
Consider an electrically neutral black hole of mass $M$ and angular momentum $J=aM$. Linear, massless field perturbations of the black hole are described by Teukolsky's equation \cite{Teukolsky72}.
Decomposing the perturbation's wave-function into radial and angular parts, ${_s\psi_{lm}}(x)=e^{i(m\phi- \omega t)} {_sS_{lm}}(\mu\equiv\cos\theta) {_sR_{lm}}(r)$, leads to separate angular and radial equations linked by a coupling constant $_sA_{lm}$, which is closely related to the Carter constant \cite{Carter68}, as we argue below.
Here, $x=(t,r,\theta,\phi)$ are Boyer-Lindquist coordinates, and $l,m$ are angular, azimuthal harmonic indices.
The spin-weight parameter $s$ gives the spin of the field, specializing the
analysis to gravitational ($s=-2$), electromagnetic ($s=-1$), scalar
($s=0$), or two-component fermion ($s=-1/2$) fields.
We use Planck units where $G=c=k_B=\hbar=1$, and shall henceforth omit the indices $s,l,m$ when possible.

Teukolsky's angular equation is \cite{Teukolsky72}
\begin{equation} \label{eq:TeukAng}
\frac{d}{d\mu} \! \left[ \! (1 \! - \! \mu^2) \! \frac{d\myS}{d\mu} \! \right] \! = \! \left[ \!
\frac{(m \! + \! s\mu)^2}{1-\mu^2} \! - \! (a \omega \mu)^2 \! + \! 2a\omega s \mu \! - \! s \! - \! A \! \right] \! \myS \! \fin
\end{equation}
The regularity of $\myS$ at the poles $\mu=\pm1$ picks out a discrete set of solutions, known as spin-weighted spheroidal wave functions
\cite[][and references therein]{BertiAlm} and corresponding eigenvalues $A(\omega)$. 
For $s=0$, these functions reduce to the familiar spheroidal wave functions \cite{Flammer}.

The radial equation can be written in the form \cite{KeshetHod07}
\begin{equation}
\label{eq:TeukRad} \left( \frac{d^2}{d r^2} +
\omega^2V^2 \right) \left( \Delta^{\frac{s+1}{2}} \myR \right) = 0 \coma
\end{equation}
where $\Delta\equiv r^2-2Mr+a^2$ vanishes at both horizons, $r_\pm\equiv M\pm (M^2-a^2)^{1/2}$. 
Here we defined $(\omega V)^2 \equiv p_r^2+sp_s^2$,
\begin{equation} \label{eq:pr}
p_r^2 \! \equiv \! \frac{[(r^2 \! + \! a^2)^2 \! - \! a^2\Delta]\omega^2 \! - \! 4Mar\omega m \! - \! (\Delta \! - \! a^2)m^2 \! - \! q\Delta}{\Delta^2} \coma
\end{equation}
$(p_s\Delta)^2 \equiv 2i[r\Delta-M(r^2-a^2)]\omega-(M-r)[2iam+s(M-r)]$,
and $q \equiv A+s-m^2$.
The QNMs are solutions of Eqs.~(\ref{eq:TeukAng}) and (\ref{eq:TeukRad}) for physical boundary conditions of purely outgoing waves at both spatial infinity and the event horizon, i.e. crossing into the black hole.
Equivalently, these solutions correspond to transmission and reflection by the black hole when the incident wave is negligible.

For given black hole ($M,J$) and field ($s,l,m$) parameters, the above constraints pick an infinite, discrete set of QNM solutions.
These modes are labelled by an index $n\in \mathbb{N}$, and are specified by their complex frequency $\omega=\omega_R+i\omega_I$ and coupling constant $A$, where $\omega_I<0$ (time decay) diverges as $n\rightarrow \infty$.
They show a complex-conjugate pairing symmetry, where for each $\{n,l,m\}$-QNM there is an $\{n,l,-m\}$-mode that satisfies \cite{Leaver85} \begin{equation}
\omega_{n,l,m}=-\omega_{n,l,-m}^* \lrgspc  \mbox{and} \lrgspc  A_{n,l,m}=A_{n,l,-m}^* \fin
\end{equation}
We thus consider only $m\geq 0$, so each mode is unique.

In addition to the pairing symmetry, the rotation of the black hole introduces a Zeeman-like splitting of the QNMs, such that for given parameters $\{n,l,m>0\}$ there are actually two modes, each belonging to a distinct QNM branch \cite{Leaver85}, which we label $+$ (new branch) and $-$ (known branch).
Each such QNM pair coalesces into a single Schwarzschild mode as $a\to 0$.

\emph{Series solution.---}
For purely imaginary large $\omega$, $S$ is thought to be of the prolate-type, such that \cite{Flammer}
\begin{equation} \label{eq:prolateA}
A^{(pr)} = i A_1 a\omega+A_0+O(|a\omega|^{-1}) \coma
\end{equation}
where $A_1=2L+1$, $L\equiv l-\mbox{max}(|m|,|s|)$ \cite{BertiAlm}, and $A_0$ is a constant; $A_0=m^2-L(L+1)/2-3/4$ for $s=0$.
It was previously argued that this applies to all QNMs in the highly-damped regime \cite{Berti04}.
However, this assumption was found \cite{Berti04, KeshetHod07} to yield only one, equatorial \cite{KeshetNeitzke08} branch of QNMs, which therefore questions the validity of the assertion.
To search for all QNMs, here we make no assumption regarding $A$, but rather solve both the angular equation (\ref{eq:TeukAng}) and the radial equation (\ref{eq:TeukRad}) simultaneously.

QNM frequencies are usually computed by series expansions of $\myR$ and $\myS$, each leading to its own three-term recurrence relation.
These commonly-used recursions were derived in \cite{Leaver85} and will not be reproduced here; a simplified form of the radial recursion may be found in \cite{HodKeshet05}.
The QNMs are determined by requiring that both recursions simultaneously converge, leading to an infinite set of QNM parameter pairs $\{\omega_n,A_n\}$.

\begin{figure}[h]
\centerline{\epsfxsize=8cm \epsfbox{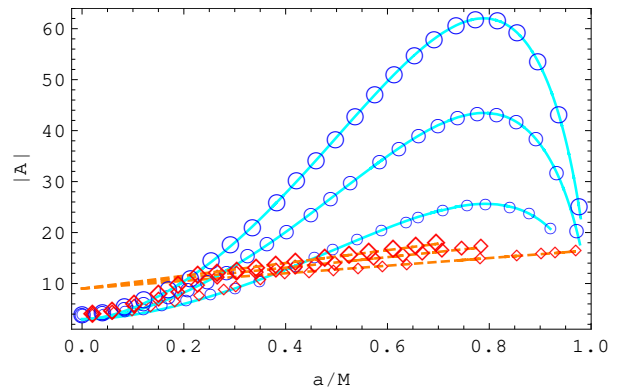}}
\caption{\label{fig:Alm}
Absolute value of the coupling constant $A$ for fundamental modes in the missing branch (circles) and in the known branch (diamonds), for $n=30$ (small symbols), $40$ (medium), and $50$ (large).
The corresponding oblate (solid curves; Eq.~\ref{eq:OblateA}) and prolate (dashed curves; Eq.~\ref{eq:prolateA} with $A_0=9$) formulae fit well above an increasingly smaller threshold $a_s(n)$.
}
\end{figure}

\begin{figure}[h]
\centerline{\epsfxsize=9cm \epsfbox{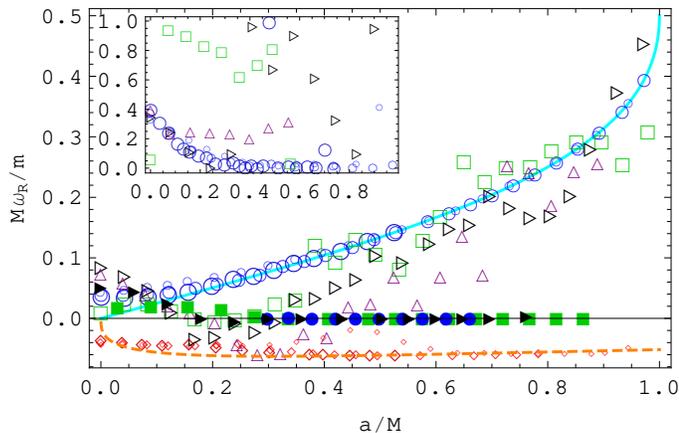}}
\caption{\label{fig:omega}
The real frequency $\omega_R$ of missing branch modes with $s=-2$ (circles), $s=-1$ (squares), $s=0$ (right triangles), and $s=-1/2$ (up triangles), for both $m>0$ (empty symbols) and $m=0$ (filled symbols).
Large symbols correspond to $n=60$.
For $s=-2$, we also show results for $n=40$ (medium symbols) and $n=20$ (small symbols), in order to illustrate the fast convergence of new branch modes onto $\omega_R=m\Omega$ (solid curve) and the slower convergence of known branch modes (diamonds) onto the known solution (dashed curve).
Inset: the fractional modulo of the imaginary frequency $|\omega_I|$ with respect to $2\pi T$ (same symbols). }
\end{figure}

In the highly-damped regime, $n\gg 1$, the series computation becomes increasingly challenging numerically.
In order to find such QNMs, including both branches, we start with pairs of easily computed \cite{Nollert99} Schwarzschild modes related by $\omega_n^+=-\omega_n^-$, defined such that $\Re(\omega_n^+)>0$ (new branch) and $\Re(\omega_n^-)<0$.
Then we gradually increase $a$ from 0 to its extremal value, in sufficiently small steps to allow our root-finding algorithm to converge.
The results are illustrated in Figures \ref{fig:Alm} and \ref{fig:omega}.

Figure \ref{fig:Alm} depicts $A$ for gravitational modes with $-s=l=m=2$ (henceforth: fundamental modes).
In the Schwarzschild limit, $A(a=0)=l(l+1)-s(s+1)$ for both branches \cite{Teukolsky72}; $A^+,A^-$ remain comparable at small values of the rotation parameter $a$.
However, beyond a certain threshold $a_s$, the two branches show markedly different behaviors.
While the known branch asymptotes to the familiar prolate scaling of Eq.~(\ref{eq:prolateA}), the new branch asymptotes to the oblate limit, given by \cite{BreuerEtAl77}
\begin{equation} \label{eq:OblateA}
A^{(ob)} = -(a\omega)^2 + 2qa\omega - (q^2-m^2+2s+1) + O(|a\omega|^{-1}) \coma
\end{equation}
where $q$ was derived in \cite{BertiAlm}, and in our case reduces to
\begin{equation} \label{eq:OblateQ}
q=\begin{cases}
1-m+2(l+s) & \text{if $m>l+2s$ ;}
\\
1+l-\mbox{mod}(l+m,2) &\text{otherwise.}
\end{cases}
\end{equation}
The split point $a_s$ becomes gradually smaller at higher $n$; comparing Eqs.~(\ref{eq:prolateA}) and (\ref{eq:OblateA}) suggests that $a_s\sim |A_1/\omega_I|$, such that $a_s\to 0$ in the highly-damped limit.

Prolate and oblate behaviors are known to interchange in the complex $\omega$ plane, along branch cuts emanating from points where different eigenvalues $A$ coalesce \cite[\eg][and references therein]{BarrowesEtAl04}.
Our results suggest that such branch cuts are present near the imaginary axis for large $|\omega|$, at least for the fundamental mode (but see \cite{BertiAlm}).
Note that both oblate and prolate asymptotic expansions are highly degenerate, $A_l^{(ob)}\simeq A_{l+1}^{(ob)}$ when $0\leq l-m+2s$ is even, and (at least when $s=0$) $A_m^{(pr)}\simeq A_{-m}^{(pr)}$.

Figure \ref{fig:omega} shows the real (figure body) and imaginary (inset) parts of the QNM frequencies of various modes.
Consider first the new branch of fundamental modes with $m>0$, depicted as empty circles of size corresponding to $n$.
It quickly converges to $\omega_R= m\Omega$ (solid curve), where $\Omega=4\pi a/\myA$ and $\myA=4\pi(r_+^2+a^2)$ are the angular velocity and surface area of the (outer) horizon.
Non-fundamental $m>0$ modes do not converge as rapidly with $n$, but are seen to approach or oscillate around this value.
All modes with $m=0$ (filled symbols) rapidly converge to $\omega_R=0$.
The inset depicts $-\omega_I/2\pi T-\lfloor -\omega_I/2\pi T\rfloor$ for various modes, where $T=(r_{+}-r_{-})/\myA$ is the Bekenstein-Hawking temperature.
It suggests that for $s\in\{-2,-1\}$, $\omega_I\to -2\pi T$ times an integer; for $s\in\{0,-1/2\}$ we cannot show such convergence.

All new branch $m>0$ modes substantially differ from the known QNM branch, for which $\omega_R$ is depicted as a dashed curve and $\omega_I$ is nearly a multiple of $2\pi T(a=0)$ \cite{KeshetHod07}.
The results suggest that the QNM frequencies of the new branch tend to the limit (solid curve in Figure \ref{fig:omega})
\begin{equation} \label{eq:QNM_freq}
\omega(n\to\infty)= m\Omega - 2\pi i T (n-s) \fin
\end{equation}
In order to derive this result, we compute the QNMs under the assumption that $A$ is given by the oblate Eq.~(\ref{eq:OblateA}).
This yields Eq.~(\ref{eq:QNM_freq}) numerically and analytically, as we show next.
However, while oblate behavior is evident for the fundamental modes in Figure \ref{fig:Alm}, numerical convergence is insufficient to establish the asymptotic angular behavior of other modes, as we discuss below.

\begin{figure}[h]
\centerline{\epsfxsize=7cm \epsfbox{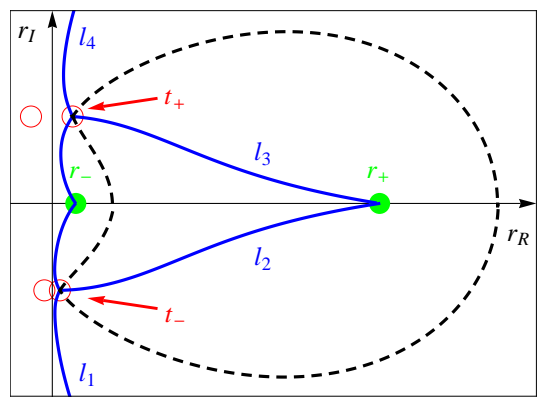}}
\caption{\label{fig:StokesNew}
Illustration of anti-Stokes (solid curves) and Stokes (dashed curves) lines in the complex $r=r_R+i r_I$ plane.
Also shown are the horizons (disks) and the turning points (circles).
Here, $a=M/2$ and $M\omega_I=-500$. }
\end{figure}

\emph{Transmission and reflection.---}
To establish the new QNM branch and study the corresponding asymptotic sector, we solve the transmission and reflection problem in the oblate, highly-damped limit.
Our analysis utilizes the Stokes phenomenon and monodromy matching, in resemblance of the prolate study in \cite{KeshetHod07, KeshetNeitzke08}.
However, the oblate geometry considered here leads to a different complexified geodesic structure, illustrated in Figure \ref{fig:StokesNew}.

We define a tortoise coordinate $z\equiv \int^r V \,dr$, such that
\begin{align} \label{eq:zGeneral}
z \sim r \text{ as } r \to \infty \, ; \lrgspc  z \to -\infty & \text{ as } r \to r_+ \coma
\end{align}
and Eq.~(\ref{eq:TeukRad}) becomes
\begin{equation}
\left( -\frac{d^2}{d z^2} + \tilde{V} - \omega^2 \right)f = 0 \coma
\end{equation}
where $f(z)\equiv V^{\frac{1}{2}}\Delta^{\frac{1+s}{2}}R$ and $\tilde{V}\equiv 3V'(r)^2/(4V^4)-V''(r)/(2V^3)$.
For an incident wave of amplitude $\nI$,
\begin{equation}
\label{eq:boundary_conditions}
f \sim
\begin{cases}
{\nI} e^{-i\omega z}+{\nR}(\omega)e^{i \omega z} & \text{as } r\rightarrow
\infty \coma\,  z\rightarrow \infty\,; \\
{\nT}(\omega)e^{-i \omega z} & \text{as } r\rightarrow r_+\coma \,
z\rightarrow -\infty\coma
\end{cases}
\end{equation}
where ${\nT}$ and ${\nR}$ are called the transmission and reflection amplitudes when ${\nI}=1$.
Here, we imposed purely outgoing boundary conditions at $r_+$, so $f$ is an eigenvector of the monodromy matrix
with eigenvalue $e^{2\pi \omega\sigma_+}$, where
\begin{equation} \label{eq:monodromy}
e^{\pm \omega\sigma_\pm} \equiv e^{\frac{\omega-m\Omega}{4\pi T}-i s/2} \fin
\end{equation}
For a discussion of the subtleties involved, see \cite{KeshetNeitzke08}.

We derive $\omega$, ${\nT}$ and ${\nR}$ in the WKB approximation, by evolving $f$ in the complex $r$-plane along anti-Stokes lines defined by $\Re(i\omega z)=0$.
These lines, identified below as complexified geodesics, are shown as solid curves in Figure \ref{fig:StokesNew} and emanate from the turning points in which $V=0$, depicted as circles.
The contour $C$ we choose is analytically continued from $r\to\infty$ to the geodesic $l_1$, continues from there to $l_2$ after avoiding the turning point $t_-$ by a $2\pi/3$ counterclockwise rotation, to $l_3$ after a clockwise rotation around $r_+$, to $l_4$ after a counterclockwise rotation around $t_+$, and finally
back to $r\to \infty$.

Monodromy matching and the mixing of WKB modes imply, in resemblance of the prolate analysis \cite{KeshetNeitzke08}, that
\begin{align}
\label{eq:matching1}
& \!\!\! {\nT}(\omega) = {\nI}(\omega)+i{\nR}(\omega); \, {\nT}(-\omega)={\nI}(-\omega)=i{\nR}(-\omega) ; \\
\label{eq:matching2}
& \!\!\! \mbox{and   }\, e^{-4\pi\omega \sigma} {\nR}(\omega) = {\nR}(\omega)+[i{\nI}(\omega)-{\nR}(\omega)]e^{-2i\omega\delta } \coma
\end{align}
where the integral in $\delta\equiv z(t_+)-z(t_-)$ is evaluated along $C$ in a single Riemann sheet
\footnote{Here, amplitudes with frequency $(-\omega)$ correspond to the potential $\tilde{V}(+\omega)$; see \cite{KeshetNeitzke08}.}.
In the highly-damped limit, $e^{-2i\omega\delta }\to 0$, so Eq.~(\ref{eq:matching2}) yields the frequency condition Eq.~(\ref{eq:QNM_freq}), as anticipated above.
Equation (\ref{eq:matching1}) guarantees flux conservation,
${\nT}(\omega){\nT}(-\omega)+{\nR}(\omega){\nR}(-\omega)={\nI}(\omega){\nI}(-\omega)$, but does not entirely fix ${\nT}(\omega)$ and ${\nR}(\omega)$.
In particular, these equations admit QNMs with ${\nI}(\omega)=0$ and total reflection modes with ${\nT}(\omega)=0$, with the same asymptotic frequencies.
For total reflection, Eq.~(\ref{eq:QNM_freq}) is precise even for finite $\omega$, as in the prolate case \cite{KeshetNeitzke08}.

\emph{Discussion.---}
After pointing out that previous studies of rotating black hole spectroscopy in the highly-damped regime were limited to only one out of two sectors, we  provide numerical evidence that the missing QNMs asymptotically approach the total-reflection branch Eq.~(\ref{eq:QNM_freq}).
This is shown in Figure \ref{fig:omega} for all spins, and is particularly striking for the converged fundamental mode.
Our results indicate that the new sector involves oblate angular wave functions with eigenvalues $A$ given by Eq.~(\ref{eq:OblateA}), at least for the fundamental mode, implying a branch cut near the imaginary $\omega$ axis.
Such behavior is seen in Figure \ref{fig:Alm}, and is consistent with our numeric and analytic computations (Eqs. \ref{eq:matching1}--\ref{eq:matching2}), which show that oblate behavior directly yields Eq.~(\ref{eq:QNM_freq}).

However, the prolate analysis \cite{KeshetHod07}, which found only one QNM branch in the highly damped limit, breaks down for the frequencies of the new branch, where the two monodromy eigenvalues in Eq.~(\ref{eq:monodromy}) become degenerate \cite{KeshetNeitzke08}.
This is not the case for our oblate analysis, where the QNMs gradually approach Eq.~(\ref{eq:QNM_freq}) as $e^{-2i\omega\delta}\to 0$.
Therefore, we cannot rule out the possibility that for some non-fundamental modes, the new branch is prolate.
Our numeric results cannot reliably resolve this issue, due to the difficulty of computing high overtones.

Consider the angular behavior of the new branch in the highly damped, oblate case. Equation (\ref{eq:OblateA}) indicates that in the Teukolsky Eq.~(\ref{eq:TeukAng}), the real part of the square brackets on the right hand side is large and negative away from the poles, so $\arg(S)$ oscillates wildly there. This suggests that $S$ is exponentially small away from the poles, which is indeed the known behavior for $s=0$ \cite[\eg][]{HunterGuerrieri82}, and appears to hold numerically for all $s$.
At high overtones, the oblate branch thus involves polar modes, in contrast to the prolate branch associated with equatorial modes; these roles are reversed at large real $\omega$.
Note that it is the new branch Eq.~(\ref{eq:QNM_freq}) which at small $n$ approaches superradiance in the extremal limit.

Consider the geodesic motion of a massless particle with angular momentum $p_\phi=m$,
complex energy $E=\omega$, and Carter's (fourth) constant of motion \cite{Carter68} fixed at some value $q_C$.
The covariant radial momentum is then given \cite{Bekenstein73} precisely by $p_r$ in Eq.~(\ref{eq:pr}), provided that $q_c=q$.
Based on both prolate and oblate results, we thus propose that Carter's constant is to be identified with $A$, up to an energy-independent term.
The potential $\omega V$ can be understood as a generalized momentum $p$ which is the quadratic mean of $p_r$ and a spin term $sp_s$.
This justifies the geodesic interpretation of Figure \ref{fig:StokesNew}; for a more detailed discussion, see \cite{KeshetNeitzke08}.
As in the prolate case, the QNM condition (\ref{eq:QNM_freq}) can now be written as a Bohr-Sommerfeld equation, $2\oint_C p \,dr= h n$, where $h$ is Planck's constant.

The real part of the new QNM branch agrees with the first law of black hole thermodynamics, $\Delta M=T \Delta \myEnt+\Omega \Delta J$, if the black hole undergoes an angular momentum change $\Delta J=m$ and an entropy change $\Delta \myEnt=0$ \cite{BertiEtAl03}.
In this sense, the new QNMs admit a simple interpretation as reversible angular momentum transitions of the black hole.
Modes with $m=0$ entail no change in $M$ or $J$; such $\omega_R=0$ modes could correspond to forbidden quantum transitions \cite{HodKeshet06}.
Interestingly, while attempts to infer the area spectrum from the QNM level spacing \cite[\eg][]{Maggiore08} yield Bekenstein's \cite{Bekenstein74} result $\Delta \myA=8\pi$ only in the slow rotation limit using the \emph{known} QNM branch \cite{Myung10}, the new branch gives $8\pi$ for arbitrary $a$.

In hindsight, several clues to the nature of the missing branch may be found in the literature.
The behavior $\omega_R=2\Omega$ was noticed \cite{BertiKokkotas03} for the fundamental mode, and associated with reversible black hole transitions \cite{BertiEtAl03}, but this was considered an intermediate or unusual case, and was later contradicted \cite[\eg][]{Berti04, BertiAlm}.
A result similar to Eq.~(\ref{eq:QNM_freq}) was postulated in Ref. \cite{Hod03} based on a corresponding $\sim|\omega|^{-1}$ drop in the recursion coefficients in the oblate limit, which may in principle truncate the expansion series.
Finally, an expression reminiscent of Eq.~(\ref{eq:QNM_freq}) was found at intermediate asymptotic frequencies in \cite{HodKeshet05}.

We thank S. Hod for encouragement and helpful discussions and suggestions.

\end{document}